# Identifying the nature of interaction between LiBH$_4$ and two-dimensional substrates: DFT study with van der Waals correction

Sung Hoon Hwang, Young-Su Lee, and Young Whan Cho[*]

High Temperature Energy Materials Research Center, Korea Institute of Science and Technology, Seoul 136-791, Republic of Korea

**Abstract**

Addressing the nature of interaction at the LiBH$_4$–carbon interface is the key to unveiling mechanism for the carbon-facilitated desorption of lithium borohydride (LiBH$_4$). Density functional theory calculations, taking into account the long range dispersion forces, have been performed to explore the interaction between LiBH$_4$, in the form of either a monomer unit or a crystalline bulk, and two-dimensional (2D) substrate, represented by graphene and hexagonal boron nitride. At the monomer–2D contact, the permanent dipole of LiBH$_4$ induces polarization of the $\pi$ electrons of the 2D, and the resultant permanent dipole – induced dipole attraction becomes the main source of binding. At the bulk–2D interface, van der Waals attraction dominates the interfacial binding rather than the dipole–dipole attraction. The absolute values of the calculated interface energy match closely with the surface energy of pristine (001) LiBH$_4$, hinting that the energy released by the formation of the interface has enough magnitude to overcome the surface energy of LiBH$_4$.

# 1. Introduction

LiBH$_4$ is a promising hydrogen storage material due to its unusually high gravimetric (18.5 wt%) and volumetric density (121 kg H$_2$ m$^{-3}$). On the other hand, its high desorption temperature and the extreme conditions to achieve reversibility provide major obstacles in materializing its practical applications [1, 2]. Recently, carbon nano-materials have been widely used to overcome those hurdles [3-6]. It has been repeatedly shown that infiltration into carbon scaffolds reduces the onset desorption temperature of LiBH$_4$ by more than 100 °C. The positive effects exerted by carbon on desorption of LiBH$_4$ are known to stem primarily from nano-confinement effects, which decrease the size of nano-particles to achieve faster diffusion and ultimately, faster rate of solid-state reaction [3, 7]. However, recent experimental reports including those where improvements are acquired from simply mixing carbon source with NaAlH$_4$ [8-10], have prompted a great deal of attention to the possibility of its pure catalyzing effects.

In the theoretical side, a number of mechanisms have been suggested for the role of carbon in facilitating the dehydrogenation of complex hydrides. It can either destabilize the borohydride (or alanate) itself or stabilize the dehydrogenated product. Berseth et al. [11] and Qian et al. [6] showed that high electron affinity of carbon nanomaterials directly affects the charge transfer between Na and AlH$_4$ unit, destabilizing the covalent Al–H bonds or stabilizing the dehydrogenated AlH$_3$. Similarly, Scheicher et al. [12] suggested the possibility of more stable daughter compound of complex hydride, LiBH$_{4-n}$ (n = 1, 2, 3) through the bond formation with carbon of C$_{60}$. In a different perspective, Xu and Ge [13] demonstrated that defects or dopants on graphene can promote H$_2$ dissociation from NaAlH$_4$. Nevertheless, the aforementioned theoretical work adopted a single or small cluster of NaAlH$_4$ or LiBH$_4$ unit to represent the bulk crystal. Such a simplified approach however, may not be sufficient in delineating realistic intermolecular interactions within the bulk crystal. Also, experimental study on bulk LiBH$_4$–carbon interface structure and their binding in the atomistic level is still in its early stage [14-16]. Even the fundamental aspect of melt-infiltration process of LiBH$_4$ into pores of the carbon scaffold [3, 14, 17] has not been thoroughly investigated. More research in the area is essential to elucidate as yet unclear mechanism for carbon-assisted dehydrogenation of LiBH$_4$.

In this work, to better understand the nature of interactions between LiBH$_4$ and carbon, we perform density functional theory (DFT) calculation on the interface structure between LiBH$_4$ and 2D surface. For more accurate description of binding, we take the van der Waals interaction into account. We adopt two-dimensional, planar graphene sheet as a carbon surface and compare its characteristic with that of isoelectronic hexagonal boron nitride (*h*-BN). First, we examine the interaction between a single LiBH$_4$ and the substrate. Interaction with Li adatom is also discussed. Then the investigation is extended to the simulation of an interface between each substrate and the low-index surfaces of the orthorhombic LiBH$_4$ crystal. By taking into account the cooperative interactions between the LiBH$_4$ units in a bulk crystal, our study would add another dimension to the recent theoretical studies focused on a monomer unit.

## 2. Computational methods

Density functional theory calculations were performed with the Vienna Ab-initio Simulation Package (VASP) [18, 19]. Exchange-correlation energy was calculated using generalized gradient approximation, in the form of the Perdew-Burke-Ernzerhof (PBE) functional [20]. Projector augmented wave potentials [21, 22] were used with a kinetic energy cutoff of 500 eV. Spin polarization was included for the investigation of Li-substrate system. To treat long range dispersion forces, two flavors of non-local van der Waals (vdW) density functionals (DFs), vdW-DF [23] and vdW-DF2 [24], were used as implemented [25, 26] in the VASP. In the case of vdW-DF, optPBE functional [27] was used instead of the original revPBE [28] since the original revPBE tends to underestimate van der Waals energy and overestimate equilibrium distance [23, 24, 29, 30].

For LiBH$_4$, orthorhombic crystal structure was adopted, and Γ-centered 6×10×6 k-point grid was used. For the hexagonal unit cell of graphene and monolayer *h*-BN, Γ-centered 30×30×1 and 18×18×1 k-point grid were used respectively, and Methfessel-Paxton smearing [31] of 0.2 eV was used in the case of graphene. The same density of k-point grid was used for the supercell calculations. Geometric relaxation was performed using the conjugate-gradient method, until Hellman-Feynmann force on each atom becomes less than 0.01 eV/Å. To investigate LiBH$_4$–2D layer interaction, 6×6 supercell of the 2D layer (72 atoms) with a length of 20 Å was used to eliminate spurious interactions between periodic images. The calculated lattice parameters for graphene and *h*-BN are 2.468 Å and 2.515 Å for PBE, 2.472 Å and 2.516 Å for vdW-DF, and 2.477 Å and 2.520 Å for vdW-DF2, respectively.

## 3. Results and discussion

### 3.1 Monomer – 2D layer interface

#### 3.1.1 Geometry and binding energy

Before we discuss the interaction between LiBH$_4$ and 2D substrates, we start by investigating the interaction between Li adatom and 2D monolayer of graphene and *h*-BN. As illustrated in Fig. 1, Li adatom occupies the central position of the hollow site on both graphene and *h*-BN. The equilibrium distance from the graphene layer is 1.71 Å. Such a short distance indicates a charge transfer from Li to the graphene [32, 33] and the resultant strong binding. In contrast, the adatom stays 3.59 Å away from the *h*-BN layer, indicating very weak adsorption. We calculate binding energy, $E_b$, using the following equation:

$$E_b = E_{system} - E_{adsorbate} - E_{layer} \qquad (1)$$

where $E_{system}$, $E_{adsorbate}$, and $E_{layer}$ represent the total energy of the hybrid system, the adsorbate (here the Li atom) and the 2D substrate layer, respectively. As expected from the equilibrium distance, binding energy for Li–graphene is significant, being –1.145 eV, while that for Li–*h*-BN is as small as –0.024 eV.

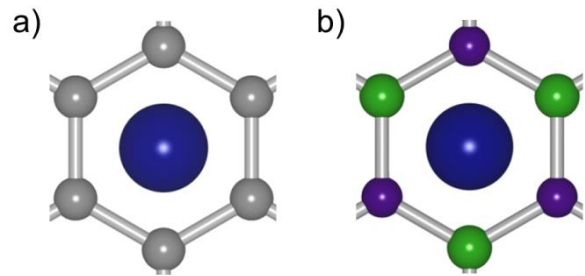

**Figure 1.** Top view of the most stable configuration of the Li on a) graphene and b) *h*-BN. Li in blue, H in pink, B in green, C in grey, N in purple.

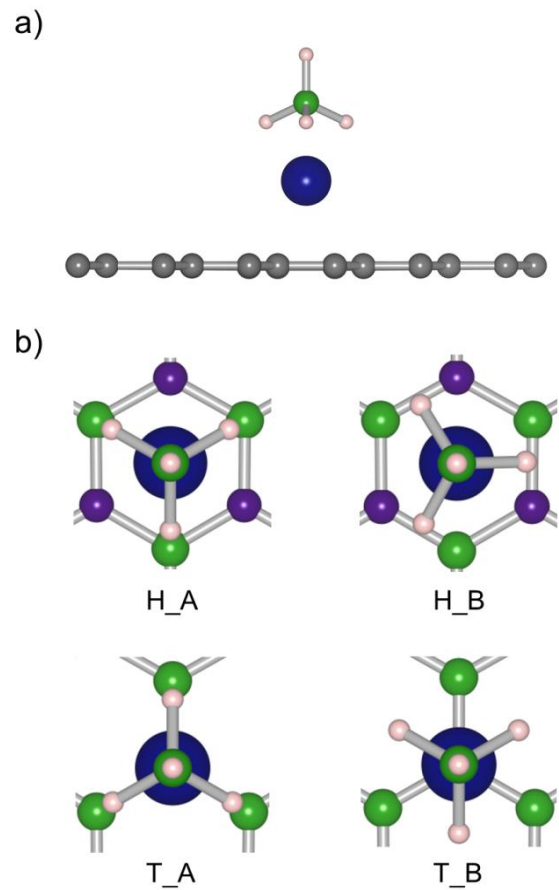

**Figure 2.** a) Side view of the equilibrium configuration of $LiBH_4$ on graphene (or *h*-BN) b) Top view of the four adsorption sites of the $LiBH_4$ on graphene (or *h*-BN)

We are interested in whether the presence of $BH_4$ unit, which competes with graphene for the valence electron of Li, would change the interaction between Li and the 2D layer. The most pronounced change in geometry due to the $BH_4$ unit, is that Li of $LiBH_4$ monomer moves away from graphene while it is attracted toward $h$-BN, and finally is located ~2.2 Å above from the both layers. In addition, $LiBH_4$–graphene and $LiBH_4$–$h$-BN share two common features in their equilibrium configurations as illustrated in Fig. 2a. Firstly, the Li occupies the position between boron of the $BH_4$ unit and the substrate layer. Reversing the position of Li and $BH_4$, such that $BH_4$ is closer to the substrate, causes significant leap in energy. Secondly, the $BH_4$ unit adopts the tridentate orientation, with three of its hydrogen atoms pointing towards the Li. This orientation is found most stable in the isolated $LiBH_4$ monomer and the presence of 2D substrate does not affect the result. It is 0.15 eV lower in energy than the bidentate orientation, reported as the equilibrium structure for $NaAlH_4$–carbon [11].

Discussing the specific position on the 2D substrate, three representative adsorption sites exist on 2D layer of graphene and $h$-BN: hollow, bridge, and top (two top sites for the heteroatomic $h$-BN). We exclude the bridge site (Li on top of C–C or B–N bond) and the top site where Li is on top of B atom in $h$-BN since they are not energetically most favorable. Fig. 2b displays four main adsorption sites for $LiBH_4$ monomer on the 2D layers. For both the hollow and the top sites, two $BH_4$ orientations are considered. We label those four sites as H_(A, B), $T_X$_(A, B), where H and T denote the hollow and the top site, respectively, and X the atom right below Li, and finally A and B the two $BH_4$ orientations. For each adsorption site, binding energy is calculated according to Eq. (1) and the results are summarized in Table 1 together with the geometry. The most stable site on graphene is H_B, with energy difference less than 1 meV from H_A. In contrast, $LiBH_4$ on $h$-BN prefers $T_N$_A site. The binding energy is –0.297 eV on graphene and –0.338 eV on $h$-BN in their equilibrium configurations. The similar magnitude of binding energy for $LiBH_4$–graphene and $LiBH_4$–$h$-

BN manifests the possibility that LiBH$_4$ adsorbs to graphene and to *h*-BN via a similar mechanism, notwithstanding differences in their electronic structure. Looking into the geometry before and after the adsorption, there is no indication of strong interaction between LiBH$_4$ and 2D substrate: LiBH$_4$ stays almost intact showing only slight stretch of Li–B distance and very little B–H bonding distance change within 0.003 Å. Stretch of C–C bond of graphene nearest to the adsorption site is also about 0.003 Å. The change in *h*-BN is more visible: the corresponding B–N distance of *h*-BN undergoes a slightly bigger stretch of 0.02 Å, with the N directly below Li protruding upwards by 0.11 Å. The ionic interaction between Li$^+$ and negatively charged N, which is absent in LiBH$_4$–graphene, explains the comparatively stronger binding and the preference for the T$_N$ site.

**Table 1. Distances and Binding Energies of LiBH$_4$ or CH$_4$ Adsorbed on the 2D layers**[a]

|  | site | 2D–Li(C) / Li–B distance (Å) | | | binding energy (eV) | | |
| --- | --- | --- | --- | --- | --- | --- | --- |
|  |  | PBE | vdW-DF (optPBE) | vdW-DF2 | PBE | vdW-DF (optPBE) | vdW-DF2 |
| isolated |  | –/1.93 | –/1.93 | –/1.92 | – | – | – |
| grapheme | H_A | 2.16/1.96 | 2.11/1.96 | 2.15/1.95 | −0.297 | −0.529 | −0.445 |
|  | H_B | 2.16/1.96 | 2.11/1.96 | 2.15/1.95 | −0.297 | −0.529 | −0.445 |
|  | T$_C$_A | 2.36/1.95 | 2.28/1.95 | 2.31/1.94 | −0.276 | −0.490 | −0.420 |
|  | T$_C$_B | 2.35/1.95 | 2.28/1.95 | 2.30/1.94 | −0.279 | −0.494 | −0.424 |
| *h*-BN | H_A | 2.12/1.96 | 2.07/1.96 | 2.09/1.95 | −0.312 | −0.539 | −0.454 |
|  | H_B | 2.13/1.96 | 2.07/1.96 | 2.10/1.95 | −0.307 | −0.532 | −0.448 |
|  | T$_N$_A | 2.21/1.95 | 2.18/1.95 | 2.17/1.94 | −0.338 | −0.547 | −0.487 |
|  | T$_N$_B | 2.21/1.95 | 2.18/1.95 | 2.17/1.94 | −0.337 | −0.545 | −0.484 |
| CH$_4$ on grapheme | T$_C$_B | 3.94/– | 3.42/– | 3.44/– | −0.014 | −0.206 | −0.148 |

[a]Data for the four different configurations as illustrated in Fig. 2 are listed. For the top sites, the distance from 2D to Li corresponds to the difference in *z* coordinates between Li and the atom on which Li is adsorbed, and for the hollow sites it corresponds to the difference in *z* coordinates between Li and the average of the six atoms constituting the hollow site.

The calculated binding energy is rather large considering the very small rearrangement in atomic positions, especially in LiBH$_4$–graphene. More importantly, the dispersion force (or van der Waals force), which would provide the major attraction for such physisorption, is not even properly included with the PBE functional. From this result, we can presume that the main source of interaction in this case comes from the dipole of $LiBH_4$, which will be elaborated in the next section. Here for better description of the binding of $LiBH_4$ on 2D, dispersion interaction is estimated using van der Waals density functionals, namely vdW-DF by Dion et al. [23] and vdW-DF2 by Lee et al. [24]. For the exchange energy in vdW-DF, optPBE developed by Klimeš et al. [27] is used instead of the original revPBE. In the recent study by Graziano et al. [34], the experimental interlayer energy of graphite is well bounded by vdW-DF2 and vdW-DF with optPBE. For further assessment of the performance of the two functionals, we test the case of $CH_4$ adsorption on graphene. The methane molecule is placed on the most stable $T_C\_B$ position [35]. Both methods give reasonable estimate of $CH_4$–graphene distance and show much improvement upon the result from the PBE functional [35]. The vdW-DF2 binding energy is very close to the experimental values of −0.12 ~ −0.14 eV [36], and vdW-DF with optPBE overestimates binding energy. Binding energies and geometric properties calculated using two different types of vdW-DFs are summarized in Table 1. Inclusion of van der Waals energy to treat $LiBH_4$–2D systems decreases $E_b$ (i.e. stronger binding) and shortens the Li–2D distance from the PBE values. However, when compared with the $CH_4$–graphene, where the dispersion interaction is dominant, the change is relatively small. If we take the vdW-DF2 values for instance, $CH_4$ moves towards graphene by 0.5 Å with order of magnitude change in binding energy, but decrease in Li–2D distance is as small as 0.05 Å and van der Waals attraction accounts for 30~40 % of the total $E_b$. This fact substantiates that long range dispersion is not a predominant source in binding of $LiBH_4$ on the 2D layers.

### 3.1.2 Charge analysis

In order to ascertain the nature of binding between the $LiBH_4$ monomer and the 2D substrate, we analyze the charge density of the system. The redistribution of charge density upon adsorption was calculated using the following equation.

$$\Delta\rho(\mathbf{r}) = \rho_{system}(\mathbf{r}) - \rho_{monomer}(\mathbf{r}) - \rho_{layer}(\mathbf{r}) \tag{2}$$

$\rho_{system}$ is the charge density of the hybrid system, $\rho_{monomer}$ and $\rho_{layer}$ are the charge density of the isolated monomer and the isolated 2D layer, respectively, which are placed at the same position as in the hybrid system. The integral of $\Delta\rho(\mathbf{r})$ over the $xy$ plane, $\Delta\rho_{int}(z)$, is plotted along the $z$-axis in Fig. 3. For comparison, the cases of Li–graphene and Li–$h$-BN are shown together. The plot in Fig 3a, characterized by large charge accumulation in the region between Li and graphene layer, and depletion around the Li position, confirms the direct charge transfer from Li to graphene. Charge redistribution in Li–$h$-BN is much smaller in magnitude. The plots for LiBH$_4$–2D systems in Fig 3b, comprise a succession of charge depletion and accumulation, both of which have almost identical areas under the curve. When the direction of dipole is reversed, the sequence of charge accumulation and depletion is changed as in Fig. 3c, proving that the charge redistribution on 2D layer is caused by the dipole of LiBH$_4$. In fact, the dipole moment of LiBH$_4$ itself is slightly enhanced by the induced dipole of the 2D layer: charge is depleted around Li and accumulated around BH$_4$ compared to the isolated LiBH$_4$. Thus, the permanent dipole − induced dipole interaction is the predominant source of binding energy of the LiBH$_4$ unit on both 2D layers. The similar binding energy for graphene and $h$-BN can be attributed to apparently similar polarizability along the $z$ direction as one can deduce from the almost overlapping profile of $\Delta\rho_{int}(z)$. However, the maximum $\Delta\rho_{int}(z)$ is slightly larger in $h$-BN (see Fig. 3b) due to more effective polarization at the N site right below Li$^+$, which explains tighter binding on $h$-BN together with the aforementioned ionic interaction.

Due to the periodic boundary conditions and the dominance of the dipole–dipole interaction in the system, it is possible that spurious dipole interaction has affected the calculated binding energies. Being a combination of an extended (2D layer) and an isolated (LiBH$_4$) system, it is difficult to apply dipole correction method commonly used [37]. We instead estimated the amount of error using a graphene flake composed of 42 C atoms saturated by hydrogen placed inside the same simulation cell. The binding energies of LiBH$_4$ to the graphene flake with and without dipole correction agree within 5 meV. Therefore we conclude that the error involved is not significant for the given simulation cell and is largely cancelled out when calculating the binding energy.

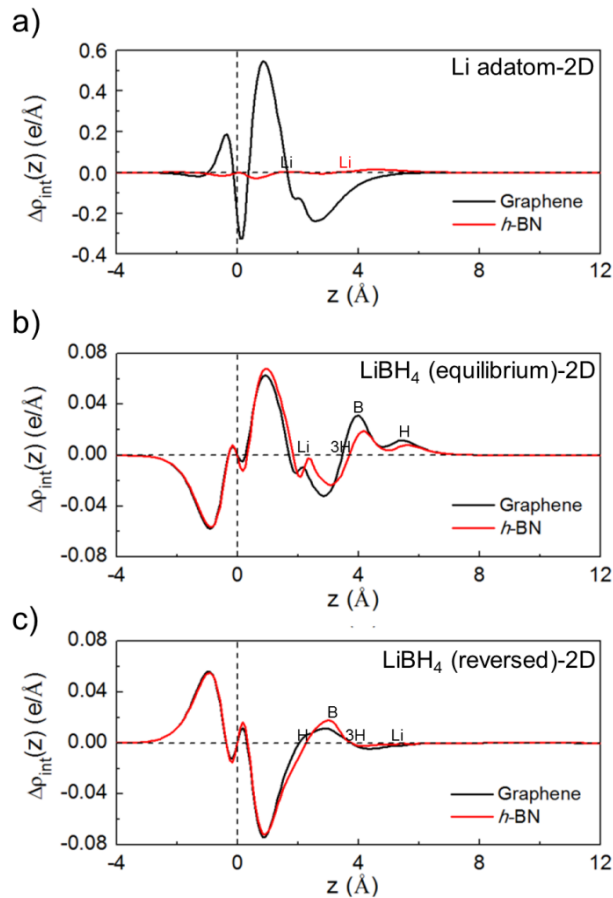

**Figure 3.** Charge redistribution in a) Li-adatom−2D, b) LiBH$_4$(equilibrium)−2D and c) LiBH$_4$(reversed)−2D. Vertical line at 0 Å indicates the position of the 2D layer. The position of each atom along the *z*-direction is marked with its atomic symbol. Distinct Li position in a) is written in black and red for graphene and *h*-BN, respectively.

**3.2 Bulk – 2D layer interface**

### 3.2.1 LiBH$_4$ Surfaces

The calculations using the LiBH$_4$ monomer have shown that the interaction with the 2D substrate largely relies on the permanent dipole–induced dipole interaction instead of a direct charge transfer. However, such interaction would not be significant unless the permanent dipole is orthogonal to the 2D plane. This fact implies that the interaction with the bulk LiBH$_4$ would probably be different since the substrate faces LiBH$_4$ units in various configurations. For this reason, we now switch to a more realistic description, using bulk LiBH$_4$–2D interface adopting orthorhombic LiBH$_4$ crystal ($o$-LiBH$_4$). Our calculated lattice parameters of $o$-LiBH$_4$ are $a = 7.253$ Å, $b = 4.382$ Å, $c = 6.628$ Å, showing 2~3% deviation in volume from the experimental values [1, 38]. Interface can be constructed by matching them with the appropriate supercell size of the 2D layer. As shown in Fig. 4, three repeated units of the orthorhombic 4-atom unit cell of the 2D layer generates similar area to that of the (001) plane of $o$-LiBH$_4$, and this specific configuration is chosen for simplicity. We checked the surface energy as a function of LiBH$_4$ slab thickness, and found that a slab with thickness of 2$c$ gives a converged surface energy. The length of the simulation cell along the $z$ direction is set to 32 Å allowing ca. 19 Å of vacuum region to prevent undesired interactions between replicas. The surface energy, $E_{\text{surface}}$, of the pristine (001) surface was calculated as follows:

$$E_{\text{surface}} = \frac{E_{\text{slab}} - E_{\text{bulk}}}{2A}, \tag{3}$$

where $E_{\text{slab}}$ is the total energy of the slab, $E_{\text{bulk}}$ is the bulk energy of the equivalent amount, and $A$ is the cross-sectional area of the slab.

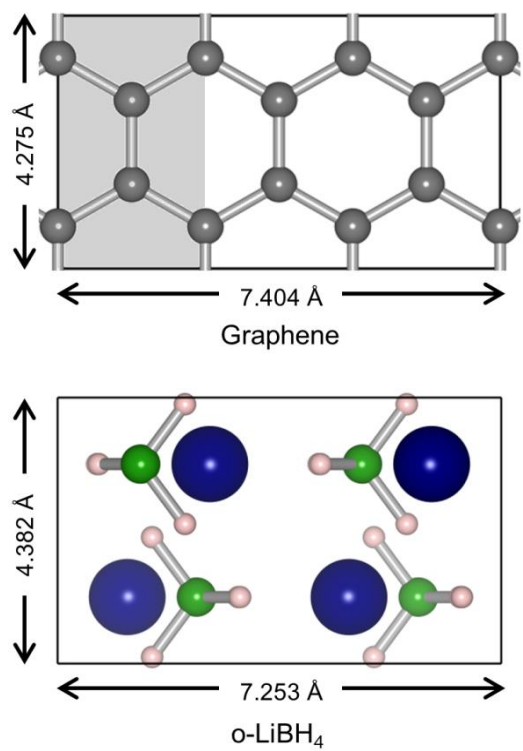

**Figure 4.** ($3a \times b$) orthorhombic graphene and ($a \times b$) (001) plane of orthorhombic LiBH$_4$. Four-atom unit cell of graphene is marked by the shaded area.

Two distinct types of termination exist for (001) as illustrated in Fig. 5. In the first type of termination, which we will denote as high-energy surface (B type), each surface Li is coordinated by two [BH$_4$]$^-$, one in bidentate and the other in tridentate orientation. Most references took this termination for the (001) surface [39-41]. When the original mirror symmetry in $xz$ plane is preserved, this termination yielded 0.337 J/m$^2$, which is in good agreement with literature values of 0.347 J/m$^2$ or 0.336 J/m$^2$ [39, 41]. Relaxation results in only slightly inward movement of the topmost Li$^+$, which increases its proximity to the surrounding [BH$_4$]$^-$ units leading to overall compression of the slab. Interestingly, breaking the mirror symmetry by displacing several atoms out of the mirror plane incurs rather dramatic reorganization of the structure as drawn in Fig. 6a. The topmost Li$^+$ is displaced by 1.2 Å to coordinate with an additional, topmost [BH$_4$]$^-$ while maintaining its proximity to the two originally neighbored [BH$_4$]$^-$ units (see Fig. 6b). The topmost [BH$_4$]$^-$ rotates to coordinate the displaced Li$^+$ in monodentate and now coordinates two topmost Li$^+$. The rotation of this [BH$_4$]$^-$ instead weakens its bonding to the Li$^+$ in the second layer from tridentate to bidentate. The more optimized Li–H coordination for the topmost layer substantially lowers the surface energy to 0.225 J/m$^2$.

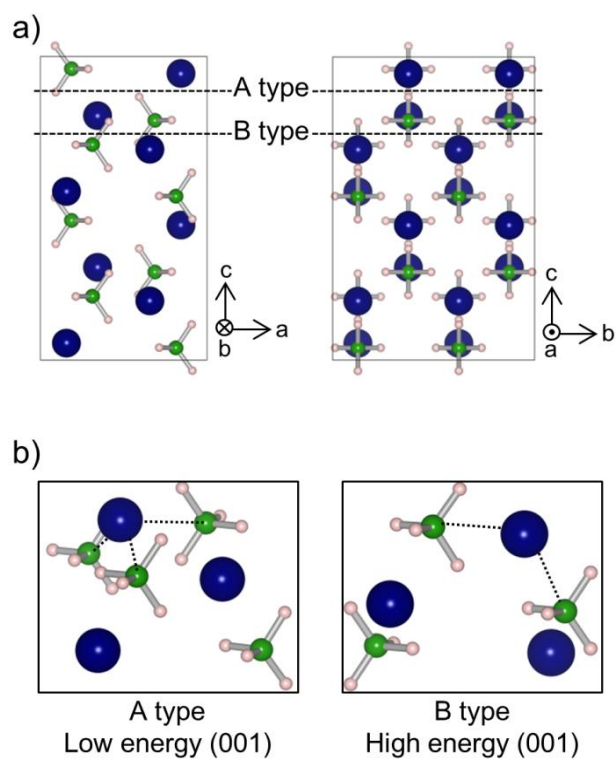

**Figure 5.** a) Bulk (1×2×2) LiBH$_4$. Two possible cleavage planes leading to the A-type and B-type (001) surface are indicated with dashed lines. b) Coordination environment of the topmost Li$^+$ of the A-type and the B-type (001) surfaces.

In the second type of termination, which we will denote as low-energy surface (A type), the topmost Li$^+$ is coordinated by three [BH$_4$]$^-$ units in the first place, all of them in bidentate orientation as shown in Fig 5. The surface energy of this termination is 0.130 J/m$^2$, significantly lower than that for the high-energy surface and even comparable to other low-energy surfaces such as (010) and (100) [39, 41]. LiBH$_4$, thus acquires higher stability when the topmost Li$^+$ is coordinated by three [BH$_4$]$^-$ units. Again when the symmetry is broken, the surface structure evolves into a new configuration in Fig. 6a. However, contrary to the high-energy surface, breaking the symmetry yields almost identical surface energy of 0.129 J/m$^2$ and much smaller relaxation effect. The most notable change is slight reorientation of [BH$_4$]$^-$ in the second layer to make shorter Li–H distance with the topmost Li$^+$.

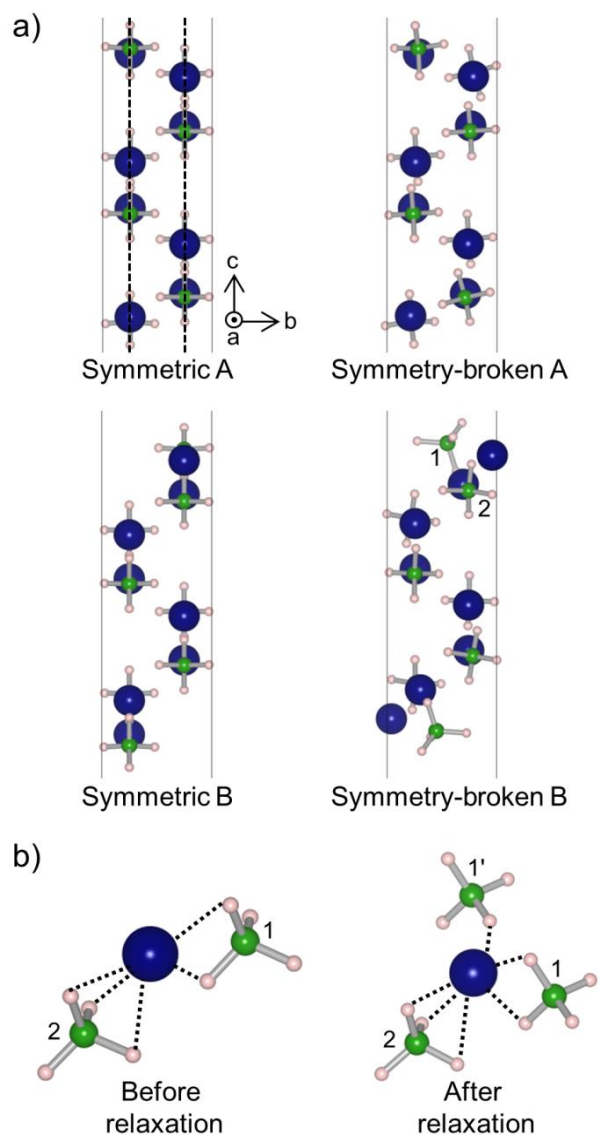

**Figure 6.** a) Relaxed structures of A- and B-type (001) surfaces with and without symmetry. Dashed lines indicate the position of the mirror planes. b) Topmost Li$^+$ and its coordinating [BH$_4$]$^-$ units in the symmetry-broken B type surface. The newly participating [BH$_4$]$^-$ after relaxation (annotated as 1′) is a periodic image of the topmost [BH$_4$]$^-$ (annotated as 1).

For modeling the bulk−2D interface, we fixed $a$ and $b$ parameters of the interface model to the in-plane dimensions of the 2D layer since graphene and $h$-BN form much stronger bonding network than LiBH$_4$. We recalculated the surface energy for the graphene lattice parameters, and the values are compared in the case of PBE in Table 2. The results are similar and the differences are within 5 % of the total value. For vdW-DFs, lattice parameters were readjusted to those acquired from each functional. Inclusion of van der Waals energy increases the surface energy by ~0.1 J/m$^2$ on average (see Table 2).

**Table 2. Surface Energies (in J/m$^2$) for the Two Types of (001) Plane using Different Methods**[a]

| method | lattice | $a$ (Å) | $b$ (Å) | A | | B | |
|---|---|---|---|---|---|---|---|
| | | | | Sym | Nosym | Sym | Nosym |
| PBE | LiBH$_4$ | 7.253 | 4.382 | 0.130 | 0.129 | 0.337 | 0.225 |
| | graphene | 7.404 | 4.275 | 0.134 | 0.134 | 0.338 | 0.217 |
| vdW-DF (optPBE) | graphene | 7.416 | 4.282 | 0.258 | 0.251 | 0.470 | 0.318 |
| vdW-DF2 | graphene | 7.431 | 4.290 | 0.218 | 0.197 | 0.429 | 0.252 |

[a]Lattice parameters are fixed to either those of LiBH$_4$ or those of graphene. A and B represent low-energy and high-energy surface, respectively. "Sym" and "Nosym" stand for the atomic configuration with and without mirror symmetry, respectively.

From the geometric point of view, vdW-DFs bring more significant relaxation effect upon the symmetry breaking and consequently yield more stable surface, i.e., difference being larger between the surfaces with and without symmetry, compared to the result from PBE. Actually the symmetric surfaces are not a local minimum and are dynamically unstable for both types of surfaces. Fig. 7 presents the minimum energy path from the symmetric to the symmetry-broken geometry obtained using the nudged elastic band method [42]. No energy is required to escape from the symmetric geometry. In the case of B type surface with PBE, another local minimum exists between the symmetric and the symmetry-broken geometry shown in Fig. 6.

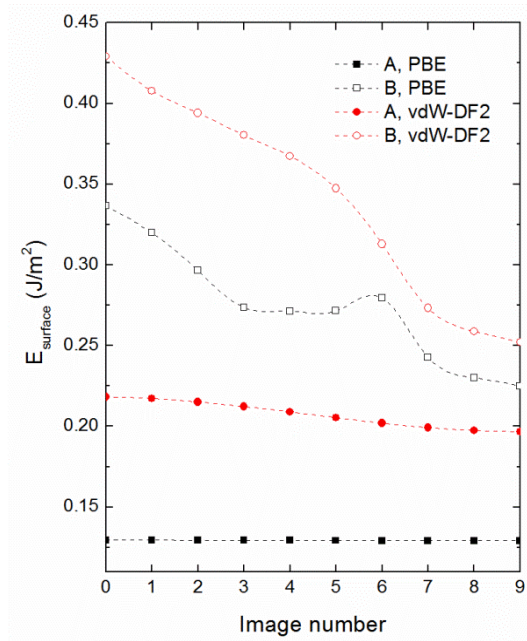

**Figure 7.** Minimum energy path from the symmetric (image number 0) to the symmetry-broken (image number 9) A- and B-type surfaces as shown in Fig. 6a. The results calculated with PBE and vdW-DF2 are plotted.

### 3.2.2 Potential energy surface of the interface

The symmetry-broken, low-energy (001) surface is used to construct bulk $LiBH_4$–2D interface. As we are ultimately interested in the degree of interaction between the bulk and the 2D, we first calculate potential energy landscape for the lateral movement of the 2D layer on top of the $LiBH_4$ surface. The calculated energy landscape would show the preferred arrangement of bulk $LiBH_4$ with respect to the 2D layers, and then we can finally assess the interface binding energy.

First, we used the fixed geometry of the $LiBH_4$ slab and the 2D layer to acquire the approximate equilibrium distance between the $LiBH_4$ surface and the 2D layer. We averaged the equilibrium distance for two arbitrarily selected configurations: one by overlapping the two lattices shown in Fig. 4 and the other by displacing the 2D layer by $b/3$ along the $y$ direction. The specific configuration is not important since the equilibrium distances from the two cases agree within 0.01 Å. At the fixed equilibrium distance, the layer was laterally moved and the landscape was obtained and is presented in Fig. 8. As one can notice from the scale, the energy difference between the energy minimum and the maximum point calculated using vdW-DF2 (slightly larger with vdW-DF) is more than ten times larger than the difference using PBE. However, the landscape does not differ much among the functionals. The distance between the topmost B atom and the $h$-BN is 4.28, 3.53, and 3.60 Å for PBE, vdW-DF, and vdW-DF2, respectively. When the distance is reduced to 3.57 Å for PBE, the energy difference becomes as large as 10 mJ/m$^2$, similar to the values from the vdW-DFs. This fact confirms that the role of van der Waals force is to reduce the overall interface distance and that the potential energy landscape is governed by the electrostatic force [43]. For instance, as illustrated in Fig. 8e, the negatively charged three topmost H atoms approach closer to the electron-rich N atoms at the energy maximum point.

One interesting feature we would like to mention is that while the absolute difference in energy between the minimum and the maximum is slightly larger in the case of $h$-BN, the energy barrier to hop between the energy minimum points is actually much lower in $h$-BN, which is somewhat counterintuitive since we expected larger friction with the heteroatomic $h$-BN.

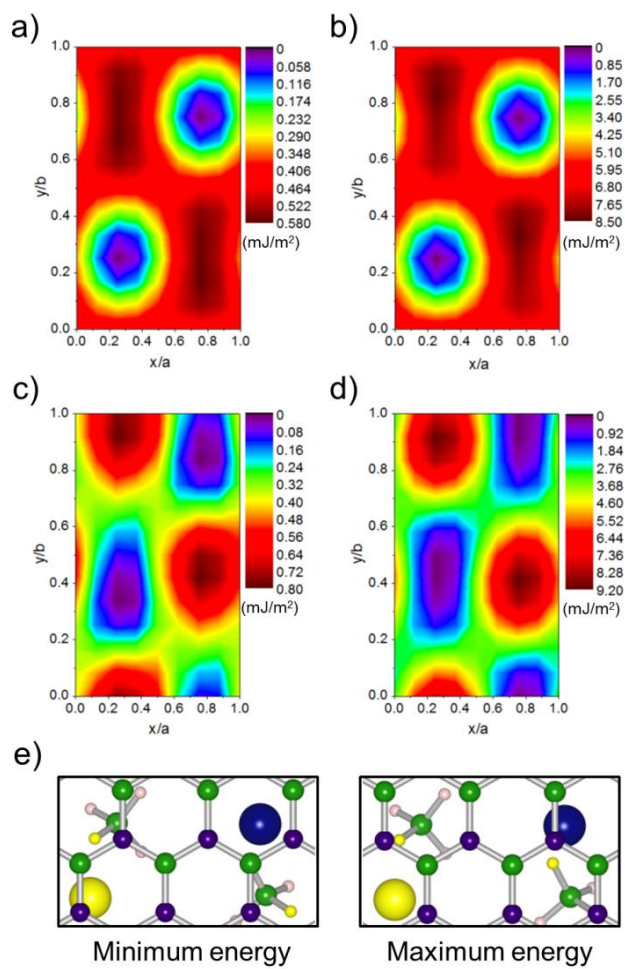

**Figure 8.** Potential energy landscapes for lateral sliding of the 2D layer: a) graphene with PBE, b) graphene with vdW-DF2, c) $h$-BN with PBE, and d) $h$-BN with vdW-DF2. The scan is limited to the unit cell of the orthorhombic graphene or $h$-BN, and $a$ and $b$ are the dimensions of such unit cell. Energy differences relative to the minimum are given in mJ/m$^2$. e) Top view of the LiBH$_4$–$h$-BN interface at the energy minimum point and at the energy maximum point (calculated with vdW-DF2). Only the atoms in the top layer are shown for clarity. The topmost Li and the three topmost H are colored in yellow.

### 3.2.3 Interface geometry and energy

Starting from the energy minimum configuration, geometry optimization was performed. The focus lies on whether the geometric relaxation would generate entirely new geometry conducive to the dipole–dipole attraction seen in the monomer–2D interface. As presented in Fig. 9, the optimization however, does not lead to significant reorganization of atoms from the pristine LiBH$_4$ slab. The most notable change is an inward movement of the topmost LiBH$_4$ by 0.1~0.3 Å, the degree being larger for B and H than Li; the B–H bond distance change is less than 0.004 Å. From the result so far, we conclude that a mere contact with ideal 2D graphene or *h*-BN would not bring a significant thermodynamic destabilization of LiBH$_4$ itself. For more exhaustive investigation on the destabilization effect by carbon scaffolds, interaction between decomposed fragments of LiBH$_4$ and graphene [12] or initiation of decomposition by more active surface sites such as bond-broken $sp^3$ carbon and functional groups is worth attention, as investigated in the case of NaAlH$_4$ [13], but is not a scope of the current work.

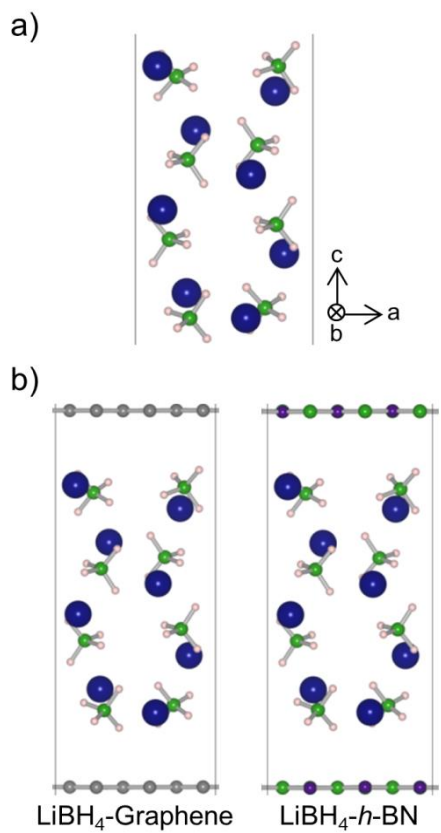

**Figure 9.** a) Relaxed structure of the A-type (001) slab, calculated with vdW-DF2. b) Relaxed structure of LiBH$_4$–graphene and LiBH$_4$–*h*-BN interface, calculated with vdW-DF2.

In different perspective, the destabilization effect of 2D surface may stem from its ability to create interface, which is composed of less stable LiBH$_4$ due to an incomplete coordination environment, instead of directly affecting the bonding in LiBH$_4$ through charge transfer. In this sense, the interface energy is very important since interface will be easily created when the interface is energetically stable. The interface energy, $E_{\text{interface}}$, is obtained from the following equation:

$$E_{\text{interface}} = \frac{E_{\text{system}} - E_{\text{slab}} - E_{\text{layer}}}{2A}, \qquad (4)$$

where $E_{\text{system}}$, $E_{\text{slab}}$, and $E_{\text{layer}}$ are the total energy of the hybrid system, the LiBH$_4$ slab, and the 2D layer, respectively, and $A$ is the cross-sectional area of the slab. The interface energies are listed in Table 3. The comparison between PBE and vdW-DFs confirms that van der Waals forces take the leading role in binding, in sharp contrast to the case of monomer−2D binding where the dipole−dipole interaction is more dominant. The surface corrugation discussed in the last section is only a fraction of the total interface energy (ca. 5 %).

To have a grasp of the magnitude of the interface energy, we make a comparison with the more familiar layered systems, i.e. graphite and $h$-BN, whose interlayer binding is governed by the dispersion force. The calculated interface energies herein can be interpreted as the energy required to exfoliate the 2D layer from the (001) surface of LiBH$_4$. Thus, we approach the interlayer binding in graphite and $h$-BN in a similar manner. We build a six-layered slab of AB-stacked graphite and AA′-stacked $h$-BN inside the simulation cell having the same $z$ dimension (32 Å) and calculate the energy required to take off the topmost layer from the surface. With vdW-DF2 functional, graphite and $h$-BN exhibit almost identical exfoliation energy of −52 meV/atom (the minus sign enters to be consistent with interface energy defined in Eq. (4).). The values agree well with the reported interlayer binding energy, which is comparable to the exfoliation energy [44], of −48, −53 and −47, −51 meV/atom for graphite and $h$-BN, respectively [34, 45]. The interface energy in Table 3 for LiBH$_4$−graphene ($h$-BN) with vdW-DF2 is converted to −32 meV/atom (−31 meV/atom), which is smaller by 20 meV/atom in absolute value than the calculated exfoliation energy for graphite or *h*-BN. This result implies that the interfacial binding is not as strong as that in graphite or *h*-BN, probably due to much looser packing of atoms at the surface layer that makes an immediate contact with the 2D layer.

**Table 3. Interface Energies between Symmetry-broken A Type Surface and 2D layer and the Distances from the Topmost Li/B/H to 2D layer**[a]

| method | interface energy (J/m$^2$) | | distance from Li/B/H (Å) | |
|---|---|---|---|---|
| | graphene | *h*-BN | graphene | *h*-BN |
| PBE | −0.007 | −0.008 | 4.54/4.32/3.41 | 4.49/4.26/3.34 |
| vdW-DF (optPBE) | −0.243 | −0.230 | 3.45/3.52/2.71 | 3.49/3.53/2.74 |
| vdW-DF2 | −0.191 | −0.180 | 3.47/3.58/2.77 | 3.48/3.59/2.80 |

[a]The distance from the topmost Li/B/H to 2D is the difference in $z$ coordinates between the corresponding atom and the average of the all the atoms constituting the 2D layer.

Still, referring to Tables 2 and 3, one can notice that the interface binding energy is so substantial as to almost compensate the surface energy. In a practical aspect, this fact indicates that the process of generating a new surface to form an interface with the 2D materials might not be costly, and it would corroborate the relative ease with which molten LiBH$_4$ fills carbon pores [3, 14], though with a caveat that our simulation deals with solid LiBH$_4$ instead of liquid. To extend the discussion further, we can link the interface energy to the exothermic event in differential scanning calorimetry data when LiBH$_4$ is melt-infiltrated into mesoporous carbon [46, 47]. The exothermic event appears right after the melting of LiBH$_4$ and has been attributed to the wetting of the pore walls [3, 14]. Now with the calculated interface energy in hand, we can roughly compare the energy change for wetting with the enthalpy for melting. Let us suppose a situation where LiBH$_4$ is squeezed into a 5-

nm wide mesopore channel bounded by two flat 2D graphene layers, like the geometry of a capacitor. The reported heat of fusion of 7.6 kJ/mol [48, 49] yields the melting enthalpy of 1.2 J/m$^2$, and the energy released by wetting is estimated to be 0.38 J/m$^2$ (two times the interface energy calculated with vdW-DF2). Direct quantitative comparison with experiment is rather difficult since, on the experimental side, melting involves partial decomposition of LiBH$_4$ or reaction with surface functional groups on mesoporous materials and, on the theoretical side, the behavior of liquid LiBH$_4$ still needs more investigation. In addition, the ratio of melting to wetting energy depends on the sizes and structures of pores. However, our result suggests that assuming a typical pore size of 5 nm, the energy released by wetting is not negligible compared to the enthalpy of melting.

## 4. Conclusions

To conclude, the density functional theory calculations have confirmed that the LiBH$_4$ monomer and the crystalline LiBH$_4$ bind to the two-dimensional substrates via different mechanisms. The LiBH$_4$ monomer induces polarization of the π electrons of 2D, leading to the dominant permanent dipole − induced dipole interaction. This accounts for rather strong binding, even in the absence of van der Waals energy terms. In contrast, the atomic configuration of the bare LiBH$_4$ slab is retained at the bulk−2D interface, and the interfacial binding mostly relies on van der Waals attraction. The role of electrostatic force is limited to determine the potential energy landscape for the lateral movement of 2D layer on the LiBH$_4$ surface. The interface energies calculated with the van der Waals density functionals have similar absolute values as the surface energies. The similarity implies that the formation of LiBH$_4$−carbon interface releases sufficient energy to overcome the surface energy holding the LiBH$_4$ units together, partly explicating the ease with which the liquid LiBH$_4$ fills the pores of carbon scaffold.

**Acknowledgement.** This work has been sponsored by the Korea Research Council of Fundamental Science and Technology and by Korea Institute of Science and Technology (Grant no. 2E24022).